\documentclass[12pt,preprint]{aastex}

\usepackage[margin=1.0 in]{geometry}
\usepackage{times}
\usepackage{amsmath}
\usepackage{rotating,tabularx}

\newcommand{\bc}{\begin{center}}
\newcommand{\ec}{\end{center}}
\newcommand{\be}{\begin{equation}}
\newcommand{\ee}{\end{equation}}
\newcommand{\bfig}{\begin{figure}}
\newcommand{\efig}{\end{figure}}

\newcommand{\Ms}{M$_\odot$}
\newcommand{\mm}{$\mu$m}

\extrafloats{100}

\received{?}
\revised{?}
\accepted{?}
	
\usepackage{amssymb}
\begin{document}
\title{High resolution Spectra of Earth-Like Planets Orbiting Red Giant Host Stars}
\author{Thea Kozakis, Lisa Kaltenegger}
\affil{Carl Sagan Institute, Cornell University, Ithaca, NY 14853}
\date{}

\begin{abstract}

In the near future we will have ground- and space-based telescopes that are designed to observe and characterize Earth-like planets. While attention is focused on exoplanets orbiting main sequence stars, more than 150 exoplanets have already been detected orbiting red giants, opening the intriguing question of what rocky worlds orbiting in the habitable zone of red giants would be like and how to characterize them. We present a high-resolution spectral database of reflection and emission spectra for nominal Earth-like planets orbiting in the red giant habitable zone from the visible to infrared (0.4-20~\mm) for planets orbiting at the Earth-equivalent distance. We also show the change of such planetary spectra through the evolution of their red giant hosts. While the luminosity of the host increases the contrast ratio between star and planet, the increased orbital distance of the habitable zone for red giant hosts relaxes the light suppression requirements close to the star, which could make such planets interesting targets to characterize and search for signs of life, if new coronagraph designs with higher suppression at larger orbital separations could be developed. We assess the feasibility of characterizing atmospheric features including biosignatures for such planets with proposed mission concept LUVOIR.
\end{abstract}

\maketitle

\section{Introduction}
Although the majority of detected exoplanets orbit main sequence (MS) stars, over 150 exoplanets have already been found around red giant (RG) stars (e.g.\ \citealt{jone14,lope16,grun17,jian18}). The characteristics of rocky planets and moons in the post-MS habitable zone (HZ) around RGs are unknown. Therefore, such places could be interesting to search for life (see e.g.\ \citealt{lore97,ster03,rami16,koza19}). Until now there have been no spectral models of such planets or moons to explore how to characterize them remotely and assess whether their spectra show biosignatures for such worlds. 

When stars exhaust their core hydrogen and leave the MS, their surfaces cool and expand, increasing their overall luminosity and engulfing their inner planetary systems. However, during this time period the orbital distance of their HZ increases past the system's original frost line, where 99.99\% of the solar system's water resides (e.g.\ \citealt{ster05}), suggesting that initially frozen rocky worlds could thaw and become partially to fully ocean-covered worlds during the RG phase of their host star (e.g.\ \citealt{lore97,ster03,rami16,koza19}). 

The post-MS HZ orbital distance changes significantly during the hydrogen shell fusion on the RG branch (RGB), becomes stable during helium core fusion on the horizontal branch (HB), and returns to rapidly changing on the asymptotic giant branch (AGB) where both hydrogen and helium shell fusion occur. During this period of stability on the HB planets can remain continuously in the post-MS RG HZ for up to 257 million years \citep{koza19}.  Although life might need longer than that time span to originate, with estimates ranging from 0.5 to 1 Gyr (e.g.\ \citealt{furn04,lope05}), subsurface life could evolve under an ice shell during the MS. Such an ice layer could melt during its host star's RG evolution and could reveal such biota through remote observations \citep{rami16}. In our own solar system, both Jupiter and Saturn will remain in the post-MS HZ for the majority of the HB, making moons such as Europa, Enceladus, and Titan interesting bodies for future temperate surface conditions in the solar system (see e.g.\ \citealt{lore97,ster03,rami16,koza19}).

Several studies explored the nature of the post-MS HZ (e.g.\ \citealt{ster03,lope05,barn13,danc13,rami16,yang17}) and the resulting planetary atmospheres \citep{koza18,koza19}, but no work had been done to quantify the spectra as well as assess whether the spectra show biosignatures for nominal Earth-like planets orbiting RG stars. Due to the wide orbital separation of the RG HZ, the spectra of such planets could be directly assessed in reflection and emission. Transit probability is significantly lower for RG HZ planets compared to planets in the HZ of MS stars due to the increased orbital distance of the RG HZ.

Assessing biosignature detectability around a wide range of host stars (for MS stars see e.g.\ reviews by \citealt{desm02,kalt07,schw18,fuji18}) 
is particularly relevant with missions on the horizon that are designed to characterize rocky planet atmospheres, namely JWST \citep{gard06}, ELT \citep{gilm07}, and future mission concepts like ARIEL \citep{tine16}, Origins \citep{batt18}, HabEx \citep{menn16}, and LUVOIR \citep{bolc16}. 

While  the orbital distance of the RG HZ increases significantly due to the higher luminosity of the host RG, increasing the apparent angular separation of such planets compared to MS HZ planets, the contrast ratio between the luminous RG hosts and their HZ planets are also smaller than for similar planets orbiting in the HZ of MS stars.  Although no mission design has yet been optimized for RG HZ planets, the relaxed inner working angle of a coronagraph could counter the increased suppression needed to adjust to the lower contrast ratio at larger angular separations. An adaptation of existing or future concepts could allow probing a new, overlooked parameter space in the search for life in the universe and answer intriguing questions on whether such planets could maintain as well as whether their spectra would show a biosphere in the RG phase of their hosts' evolution.

Note that ground-based high resolution spectroscopy has already characterized atmospheric species for unresolved planets like HD 179949 b using the planet's known motion during the observations (e.g.\ \citealt{snel13,brog14,birk18}).  We extrapolate that a similar approach could be used to counter the increased luminosity of RG hosts to observe their HZ planets, making such planets interesting targets for observations with Extremely Large Telescopes.

As a starting point to explore how to search for life around RG stellar hosts, we simulated Earth-like planets around a variety of RG spectral types \citep{koza19}.  While initially frozen worlds could harbor very different kinds of biota,  how life would evolve on such objects is yet unknown.  In this study we present spectra for nominal Earth-like planets in the RG HZ both i) at the Earth-equivalent distance, where they receive similar irradiation as modern Earth and ii) for a specific orbital distance in the RG HZ for several points throughout the RG's evolution for stellar mass tracks of 1.3, 2.3 and 3.0 \Ms. For a detailed model description see \cite{koza19}, which discusses the planetary atmospheric models and their corresponding UV surface environments. These models include planetary atmospheric erosion as well as semimajor axis evolution resulting from its host's mass loss during the RG phase \citep{koza19}. Due to the wide orbital separation of the HZ around luminous RGs, planetary atmospheric mass loss via erosion is limited to less than 10\% for planets with a 1~bar surface pressure (ibid). A summary of RG hosts and their HZ model data from \cite{koza19} is shown in Table~\ref{RG_targets}.

\begin{table}[h!]
\caption{Properties of the selected red giant hosts \label{RG_targets}}
\begin{center}
\footnotesize
\begin{tabular}{cllrrrrrrr}
\hline
Spectral & Star & Star &  Dist. & Radius & Planet & 1 AU equiv. & Mass track & $a$ for max \\
 Type & Name &  T$_{\tiny \mbox{eff}}$ (K) & (pc) & (R$_\odot$)  & T$_{\tiny \mbox{surf}}$ (K)$^a$ & dist (AU) & (\Ms) & HZ (AU)$^b$ \\
 \hline
G5 III	& HD 74772          & 5118  & 70.18     & 12.90  & 289.3 & 10.12 & 3.0 & 18.2\\ 
G8 III	& HD 148374         & 4948  & 155.28    & 14.52 & 294.5  &  10.65  & 3.0 & 18.2\\ 
K0 III	& $\beta$ Gem       & 4865  & 10.36     & 8.8   & 295.0  & 6.24  & 2.3 & 12.0\\ 
K0 III	& $\beta$ Ceti      & 4797  & 29.5      & 16.78 & 295.5  &  11.57  & 3.0 & 18.2\\ 
K2 III	& $\iota$ Draconis  & 4445  & 31.03     & 11.99 & 298.6  & 7.10  & 1.3 & 12.5 \\ 
K2 III	& $\theta$ Doradus  & 4320  & 151       & 16    & 299.1 & 8.94  & 1.3 & 12.5  \\ 
K3 III	& $\alpha$ Boo      & 4286  & 11.26     & 25.4  & 303.6  & 13.98 & 2.0 & 12.2 \\ 
K5 III	& $\gamma$ Draconis & 3989  & 47.3      & 53.4  & 304.7  & 25.45 & 2.3 & 12.0\\ 
 \hline
 \end{tabular}
 \end{center}
 {\footnotesize $^a$ Planetary surface temperatures were modeled in \cite{koza19}, whereas all other quantities in this table were derived from the host's stellar parameters\\ 
 $^b$Semimajor axis values correspond to a planet's inital orbital position resulting in the maximum continuous amount of time in the HZ of its RG host, taking stellar mass loss and planetary orbit changes into account} 
 \end{table}

Section~\ref{methods} describes our methods, Section~\ref{Earth-eq} discusses the spectra of planets in the post-MS HZ at Earth-equivalent distance as well as throughout their host's RG evolution, along with a discussion on coronagraph design improvements necessary to enable characterization of RG HZ planets, as well as LUVOIR simulations to simulate required integration time to detect spectral features if current coronagraph designs are used.  Section~\ref{conclusions} discusses and concludes our paper.

\section{Methods\label{methods}}

\subsection{Stellar Hosts and Planet Atmospheric Spectra Models}
\begin{figure}[h!]
\centering
\includegraphics{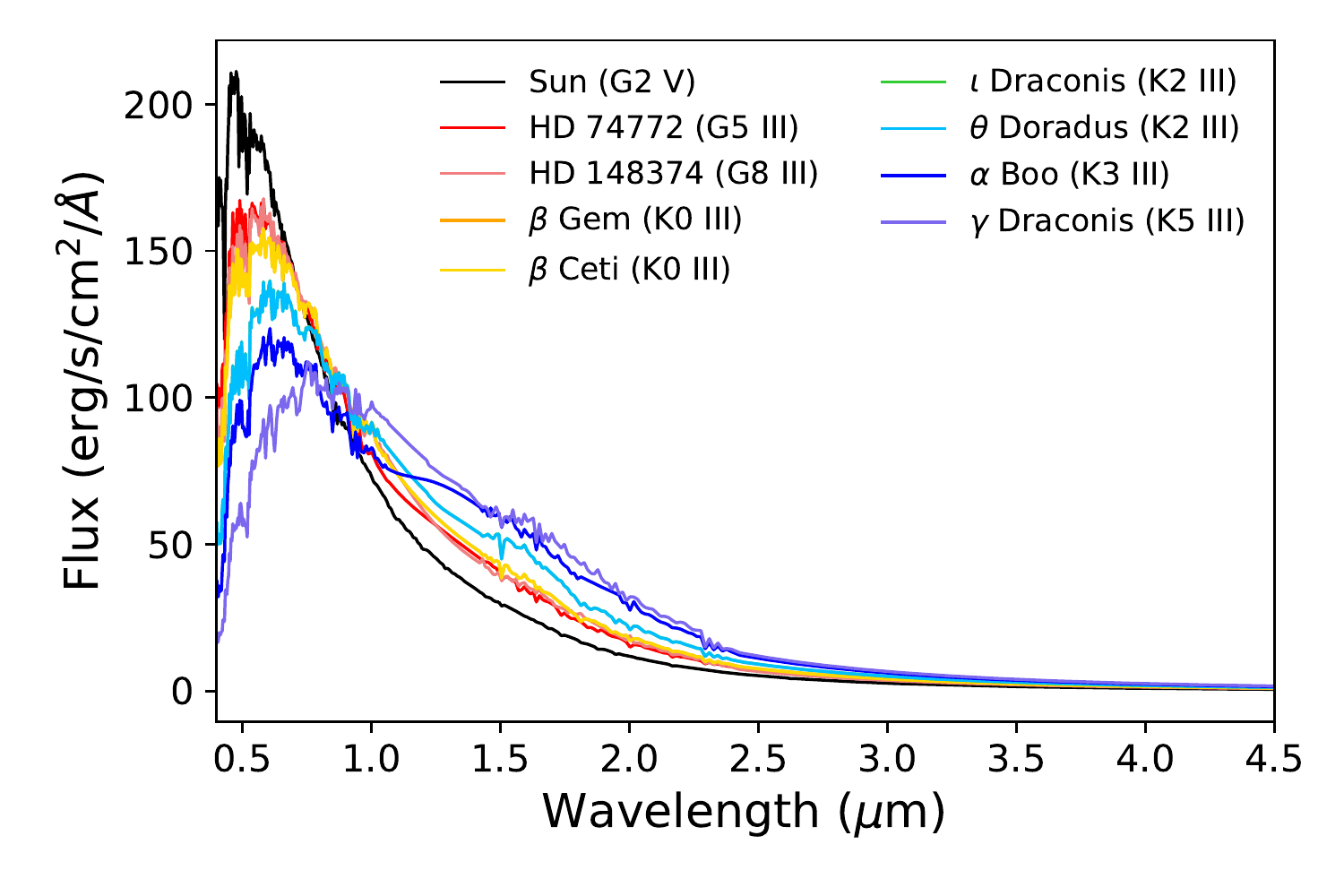}
\caption{Incident stellar spectra of the red giant hosts scaled to the 1 AU-equivalent distance, shown at a resolution of R = 140. Red giants with cooler surface temperatures show longer peak wavelengths as well as additional spectral lines due to molecular formation in their atmospheres. See Table~\ref{RG_targets} for information on individual hosts. The solar spectrum is shown for comparison in black. \label{fig:RGsinput}}
\end{figure}

We model Earth-like planet reflection and emission spectra orbiting 8 different RG host stars, which have IUE UV data\footnote{http://archive.stsci.edu/iue/}. Their incident stellar spectra, which illuminates our model planets, consist of combined IUE and Pickles Atlas \citep{pick98} luminosity class III spectra (see Table~\ref{RG_targets} and Figure~\ref{fig:RGsinput}).  We estimate evolutionary phase and mass track using results from \cite{stoc18} and H-R diagram fitting.  For a full description of these stellar spectra see \cite{koza19}. We created the planetary atmosphere models and high resolution spectra using \emph{Exo-Prime} (e.g.\ \citealt{kalt20}), a 1D coupled climate-photochemistry-radiative transfer code developed for terrestrial planets.

For each planetary atmospheric model, described in \cite{koza19}, \emph{Exo-Prime} generates a high resolution reflection and emission spectra in wavenumber steps of 0.01 cm$^{-1}$ from 0.4 to 20 \mm, using a line-by-line radiative transfer model originally developed to retrieve trace gases in Earth's atmosphere \citep{trau76} and later adapted to simulate emergent and transmission exoplanet spectra (e.g.\ \citealt{kalt07,kalt09}). For each atmospheric layer line shapes and widths are calculated individually with Doppler- and pressure-broadening with several points per line width. We include the most spectroscopically relevant molecules in our calculations: H$_2$O, CO$_2$, N$_2$O, NO$_2$, HNO$_3$, O$_3$, CH$_4$, O$_2$, CO, OH, H$_2$CO, CH$_3$Cl, and O, as well as Rayleigh scattering, using the HITRAN 2016 line lists \citep{Gordon2017}. 

We use a Lambert sphere approximation for a disk-integrated planet and model modern Earth-like surfaces for clear sky and 60\% coverage of Earth-like clouds (following \citealt{kalt07}). Our modern Earth model surface consists of about 70\% ocean, 2\% coast, and 28\% land, with the land comprised of about 30\% grass, 30\% trees, 9\% granite, 9\% basalt, 15\% snow, and 7\% sand. The 60\% coverage by modern Earth-like clouds assumes a distribution of clouds of 40\%  at 1 km, 40\%  at 6 km, and 20\% at 12 km. We are using the same wavelength dependent cloud albedo for all three cloud layers. Note that we do not change the height or properties of clouds for different host stars. Cloud feedback for different host stars is an area of ongoing research (see e.g.\ review by \citealt{hell19}). Effects of different cloud coverage and heights on the climate and spectra of Earth-like planets orbiting RG have not been explored in this study, but have been examined for planets orbiting main sequence stars with MS hosts with similar effective stellar surface temperatures (e.g.\ \citealt{fuji11,kitz11,kitz11b,zsom12,rugh13,kawa19}).  All albedo data are from the USGS Digital Spectral Library\footnote{https://www.usgs.gov/labs/spec-lab/capabilities/spectral-library} and the ASTER Spectral Library\footnote{https://speclib.jpl.nasa.gov}.

All spectra can be downloaded in high resolution (with resolution of a minimum of 100,000 at any wavelength) from the online RG planet spectra catalog\footnote{https://doi.org/10.5281/zenodo.4027344}. Note that we have not added any noise to the spectra to provide theoretical input spectra for any instrument.

For ease of comparison in the figures, we rebin the spectra to a resolution of 140 using a triangular smoothing kernel. We chose the resolution of 140 for our figures based on instruments on JWST \citep{gard06} with resolution between 100 and 150 and instrument designs for ARIEL \citep{tine16}, WFIRST \citep{gree12}, Origins \citep{batt18}, HabEX \citep{menn16}, and LUVOIR \citep{bolc16}.

\subsection{Coronagraph simulation of detectability of  spectral features}

While no existing or proposed instrument concept is optimized to observe RG HZ planets, as an example we use one of the proposed mission concepts, LUVOIR, to estimate the required integration times in the visible to observe spectral features in reflected light for RG HZ planets for i) the current design of one LUVOIR's proposed coronagraphs, and ii) a hypothetical coronagraph design, which allows for a lower contrast ratio by a factor of 100 at the wider orbital separation, corresponding to the RG HZ. 

We use a coronagraph noise simulator originally developed for WFIRST-AFTA \citep{robi16,lust19} and system parameters from the LUVOIR Final Report\footnote{https://asd.gsfc.nasa.gov/luvoir/reports/}.  In particular we use the predicted parameters of LUVOIR-A (15 meters in diameter) and the  Extreme Coronagraph for Living Planet Systems (ECLIPS), a coronagraph with imaging spectroscopy for characterizing exoplanetary atmospheres, with a proposed suppression of 10$^{10}$ between its inner working angle (IWA) of 3.5 $\lambda/D$ and its outer working angle (OWA) of 64 $\lambda/D$. We then hypothesize a coronagraph concept can be optimized for planets in the RG HZ, producing a 10$^{12}$ suppression at the wider RG HZ separation out to an OWA of 230 $\lambda/D$.

\section{Results: spectra of Earth-like planets orbiting red giant hosts \label{Earth-eq}}

\subsection{Spectra of Earth-like planets in the RG HZ receiving modern Earth irradiation}

\begin{figure}[h!]
\centering
\includegraphics[scale=0.75]{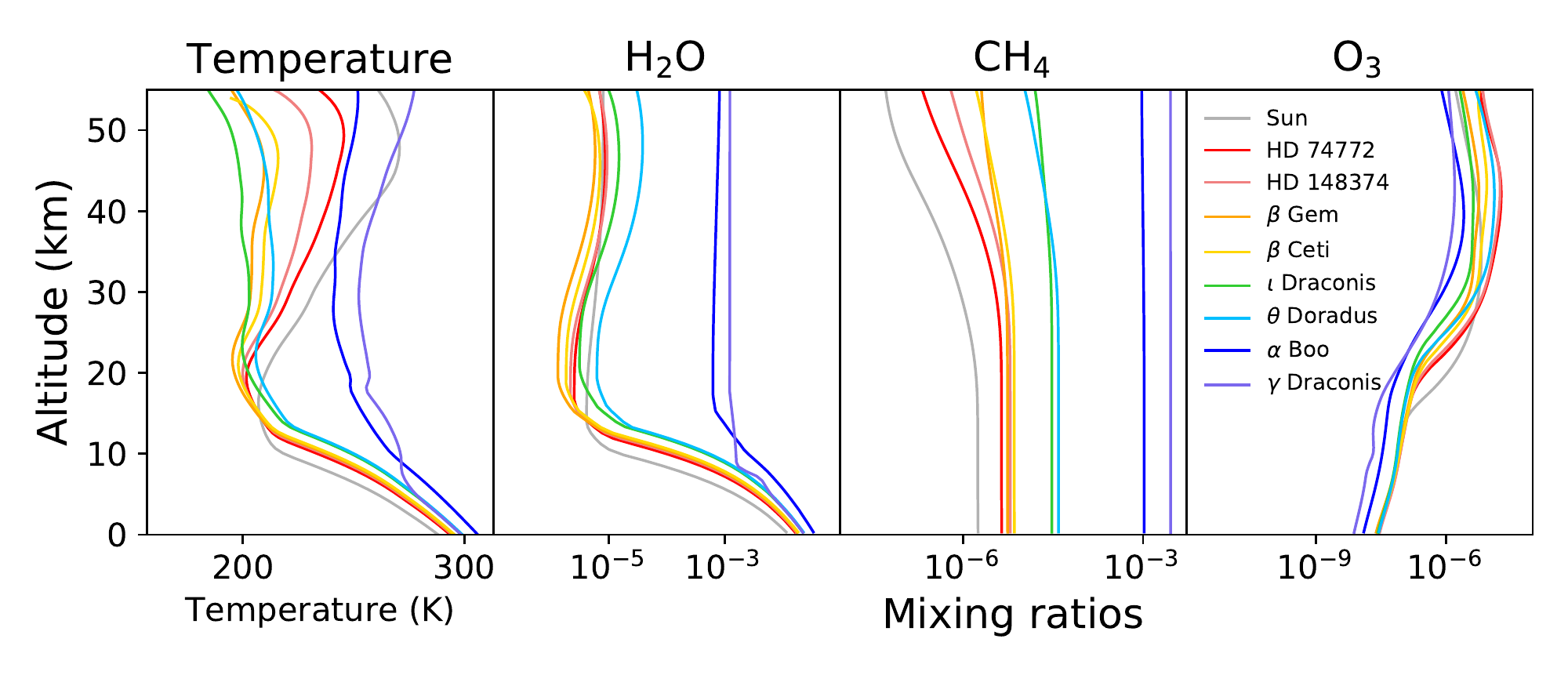}
\caption{Temperature and mixing ratios for H$_2$O, CH$_4$, and O$_3$ for nominal Earth-like planet models orbiting in the habitable zone of 8 red giants within 25pc (T$_{surf}$ 5200 K to 3600 K) at the 1 AU equivalent distance (from \citealt{koza19}).  RG hosts are ordered from hottest to coolest in the legend, with the modern Earth-Sun system in black for comparison. \label{fig:T}}
\end{figure}

Temperature, chemical mixing ratio profiles (see Figure~\ref{fig:T}), and UV surface environments for our planet models are described in detail in \cite{koza19}. We summarize the major model characteristics of that paper here to link them to the atmospheric features that are shown in the spectra and contrast ratios. 

For all models we kept surface outgassing rates at modern Earth levels for H$_2$, CH$_4$, CO, N$_2$O, and CH$_3$Cl and maintain constant mixing ratios of O$_2$ at 0.21 and CO$_2$ at 3.55$\times10^{-6}$. The surface pressure is set by the initial 1 bar pressure and consequent atmosphere erosion. 
Planetary surface temperatures increase with decreasing stellar surface temperature of their RGs hosts due to the shift in peak wavelength to longer wavelengths for cool stars because of the higher efficiency of surface heating via redder light and decreased effectiveness of Rayleigh scattering for longer wavelengths (see e.g.\ \citealt{kast93}). This increase in planetary surface temperature for cooler RG hosts increases the amounts of atmospheric H$_2$O for the warmer model planets. 

Planets orbiting hotter RGs receive higher incident UV irradiation, thus show higher ozone (O$_3$) production, and create more hydroxyl (OH), a by-product of ozone.  This leads to lower levels of methane (CH$_4$) for planets orbiting hotter RGs both due to depletion via OH and higher rates of UV photolysis.  Temperature inversion in Earth-like atmospheres is primarily caused by atmospheric heating due to ozone absorption and thus becomes more pronounced for planets orbiting hotter RGs. CH$_4$ concentrations increase in our models for cooler RG hosts, where CH$_4$ heats the upper atmosphere, further reducing temperature inversion for planets orbiting cooler RG hosts. Several studies have found similar results for planets orbiting in the HZ of MS stars with similar stellar surface temperatures (e.g.\ \citealt{segu05,rugh15}).

The planet models around the two coolest RGs hosts ($\alpha$ Boo and $\gamma$ Draconis, a K3 III and a K5 III star, respectively) have high surface temperatures and show stratospheric H$_2$O mixing ratios of more than 3$\times10^{-3}$, which was initially defined as the boundary of the classical moist greenhouse regime \citep{kast93}. However, recent work using both 1D and 3D models has shown that planet models around cool stars can often reach H$_2$O mixing ratios within this regime, yet maintain habitable surface temperatures, particularly around inactive stars (e.g.\ \citealt{kopp17,chen19}).  Although planets in the classical moist greenhouse regime have large amounts of stratospheric H$_2$O that could be susceptible to destruction via photolysis, around inactive cool late K and early M stars the stellar UV flux is low and therefore photolysis rates can remain low. While our models calculate H$_2$O photolysis rates, they do not include potential H loss as a result.

\begin{figure}[h!]
\centering
\includegraphics[scale=0.53]{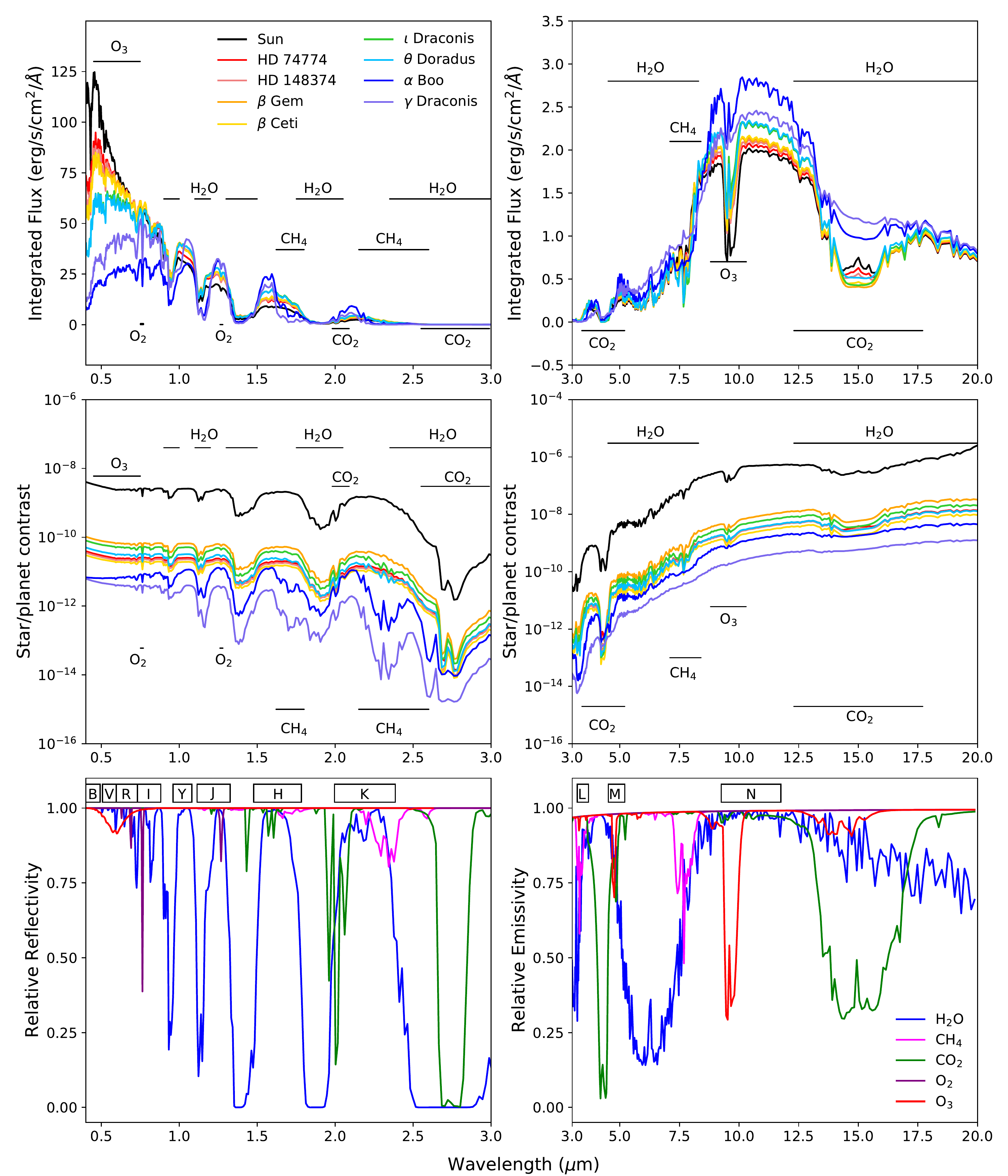}
\caption{Reflection (left) and emission (right) spectra for Earth-like planets orbiting their red giant hosts at the 1 AU equivalent distance. The top row shows the planet's flux, the middle the contrast ratio between the planet and its host star, and the bottom row the absorption of the major chemical species for the Sun-Earth case are shown individually for reference.  RGs are ordered from hottest to coolest in the legend, with the modern Earth-Sun system for comparison. \label{fig:RG_spectra}}
\end{figure}

Figure~\ref{fig:RG_spectra} shows the planets' reflection and emission spectra. Atmospheric feature depth in reflection spectra is determined by the abundance of the chemical species and the surface reflectivity for Earth-like atmospheres, while the depth of spectral features in an emission spectra is set by the abundance of the chemical and the temperature at which a chemical absorbs/emits versus the overall emitting temperature of the planet. Higher surface reflectivity results in more reflected starlight and thus deeper spectral features in the visible for the same chemical abundance and incident stellar spectra. Larger temperature contrast between the emitting/absorbing layer compared to the continuum at that wavelength give more depth to spectral features in the infrared for the same chemical abundance. The increasing peak wavelength of the emission for cooler RGs reduces the shortwave incident stellar light, the amount of reflected light and therefore the absorption depth of chemicals in reflected light at short wavelength. Planets orbiting cooler RG hosts also show decreasing temperature inversion in the model atmospheres, reducing the depth of absorption features in the infrared for planets orbiting cooler RGs. 

Figure~\ref{fig:RG_spectra} shows the spectra of nominal Earth-like planets orbiting RG hosts at the 1 AU equivalent distance, as planetary flux (top row),  contrast ratios of the planet to its host star's flux (middle row), as well as the relative absorption of the individual chemicals in the Earth-Sun model for reference, to clarify the overlap of individual spectral lines (bottom row). These individual chemical plots are generated by removing the opacity of all other molecules from the original atmosphere, and calculating only the opacity due to one particular atmospheric species (see \citealt{kalt07}). Full 0.4-20 \mm\ wavelength coverage for the reflected light and emission spectra are available online\footnote{https://doi.org/10.5281/zenodo.4027344}. Due to the host stars' effective temperatures both the effects of reflected light and planetary emission are included in the wavelength range between 2-5 \mm, where the contribution of the reflected and emitted flux change over to become the dominant source of the planetary flux. Modern Earth spectra and contrast ratio are shown in black for comparison and labeled ``Sun". 

At a resolution of R = $\lambda/\Delta\lambda$ = 140, shown in (Figure~\ref{fig:RG_spectra}), all planet spectra show features of H$_2$O, CO$_2$, CH$_4$, O$_3$ and O$_2$ (top row). Planet spectra in the visible show the lowest flux for planets orbiting cool RG hosts, although their increased surface temperature gives them the largest emission flux in the infrared. The contrast ratio between the host star and its planet generally decreases with decreasing RG surface temperature, making planets orbiting in the HZ of hotter RG hosts easier targets to observe (middle row of Figure~\ref{fig:RG_spectra}). This is due to the generally larger radii/luminosity of cooler RGs.  The contrast ratio in the visible is about 10$^3$ lower than in the infrared, similar to planets orbiting in the HZ of MS stars. The black line shows the modern Earth-Sun spectra as well as contrast ratio for comparison. While the planetary flux of the modern Earth-Sun system is similar for planets in the HZ of RG hosts, the contrast ratio of the star to planet decreases by about a factor of 15 up to 10$^3$ for planets orbiting in the HZ of RGs because of the increasing luminosity of the RG hosts, which increase with decreasing surface temperature (see Table~\ref{RG_targets}).

H$_2$O absorption bands are present at 0.9, 1.1, 1.4, 1.9, 5, and 17 \mm. The abundance of the H$_2$O absorption features increase with increasing planetary surface temperature for cooler RG hosts. In addition, cooler RG hosts provide lower stellar UV environments, decreasing water photolysis. CO$_2$ shows strong absorption features at 2 and 15 \mm. The temperature inversion in the model atmospheres decreases for cooler RG hosts, decreasing the inversion indication in the center of the CO$_2$ absorption feature at 15 \mm\ for cooler RG hosts. 

Hotter RG hosts show increased O$_3$ production due to higher stellar UV flux, which deepens the 9.6 \mm\ O$_3$ feature because of the larger temperature difference between the hotter stratosphere and the planet's overall emission temperature for these models. Note that the O$_3$ feature at 0.6 \mm\ in reflected light is very shallow for all models. Absorption features for CH$_4$ can be seen at 1.7, 2.4 and 7.7 \mm. CH$_4$ is depleted both through UV photolysis and reactions with OH (a by-product of O$_3$), causing shallower feature depth in the reflected light at 1.7 and 2.4 \mm\ for hotter RG hosts with more UV. The IR absorption feature at 7.7 \mm\ shows more variation due to the additional effect of the decreasing temperature difference between the absorption layer and the overall planet emission temperature for cooler RG host models. 

\begin{figure}[h!]
\centering
\includegraphics[scale=0.8]{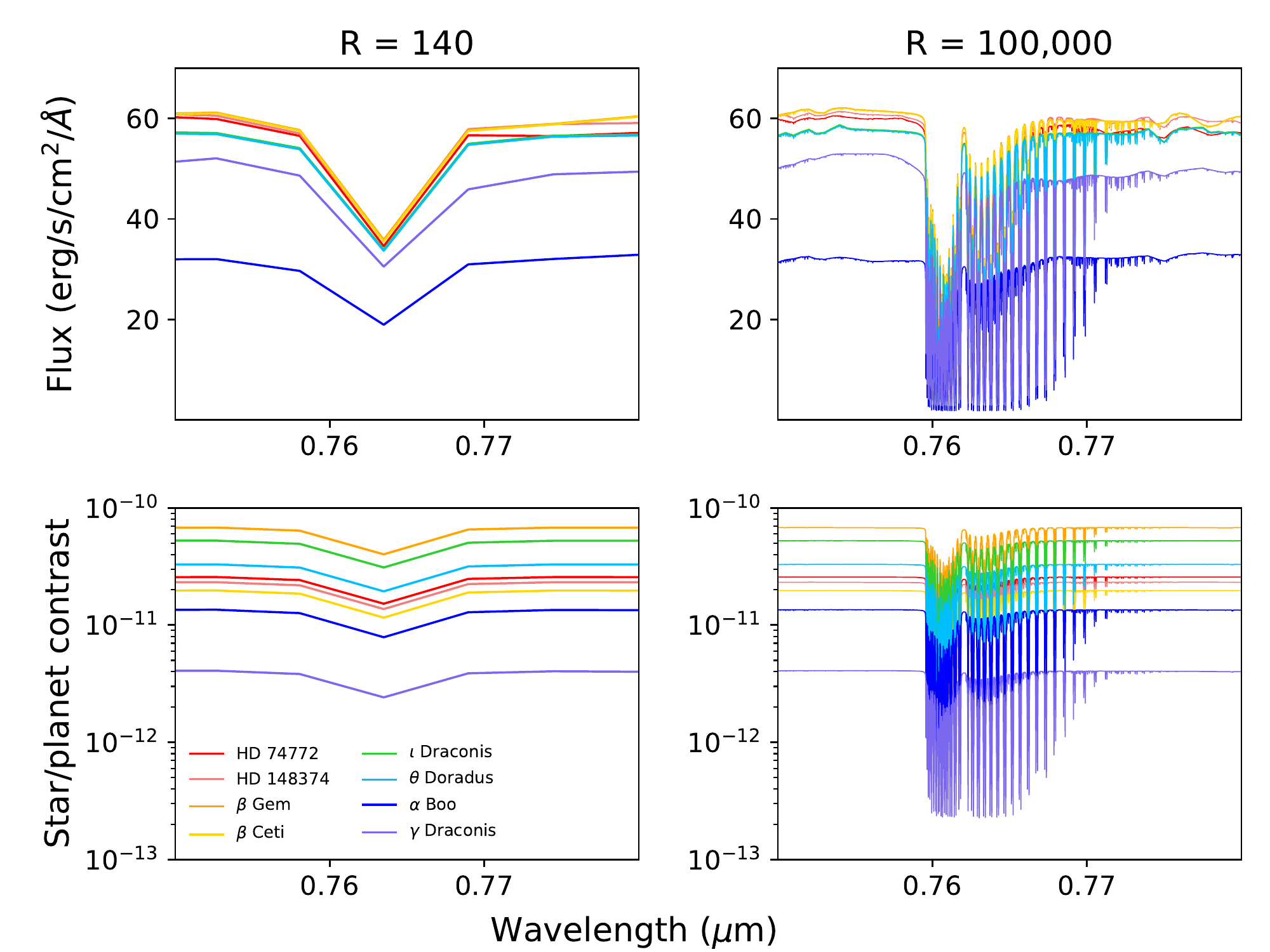}
\caption{O$_2$ features at 0.76 \mm\ as (top) planet flux and (bottom) star/planet contrast shown at a resolution of (left) 140 and (right) at high resolution of 100,000, as proposed for the HIRES instrument on the ELT. \label{fig:O2}}
\end{figure}

Absorption features for O$_2$ (or O$_3$) in combination with CH$_4$ (see e.g.\ \citealt{love65,lede65,lipp67}) can indicate life on a planet like Earth. This combination of features can be seen in Figure~\ref{fig:RG_spectra} for all planet models in the RG HZ. The depth of the O$_2$ feature at 0.76 \mm\ in the reflection spectra becomes difficult to identify at a resolution of 140 for cooler RGs because of the low incident flux level for cool RGs at these wavelengths. As an example of how such features appear in high resolution, Figure~\ref{fig:O2} shows the absorption features of O$_2$ at 0.76 \mm\ for all planet models at a resolution of R =  100,000, which is proposed for several upcoming instruments such HIRES on the ELT \citep{gilm07}. For ground-based observations Doppler shift due to the motion of the planet compared to terrestrial features can increase the detectability of spectral features in high resolution (e.g.\ \citealt{snel13,brog14,birk18}). 

\subsection{Spectra of Earth-like planets in the HZ through the red giant's evolution \label{evol_dist}}

In this section we discuss the spectra for planet models through their RG's evolution. Temperature and mixing ratios for the major chemicals for all model planets throughout their RGs host star evolution are shown and discussed in \cite{koza19}. We model high resolution spectra for these planets orbiting RG hosts for 3 mass tracks of 1.3, 2.3, and 3.0  M$_\odot$. For each mass track we model the planetary atmosphere at multiple points on the orbit that provides the planet the maximum amount of time continuously in the RG HZ for points in the RG's mass track evolution where we have RG UV data (see \citealt{koza19}).

\begin{figure}[h!]
\centering
\includegraphics[scale=0.57]{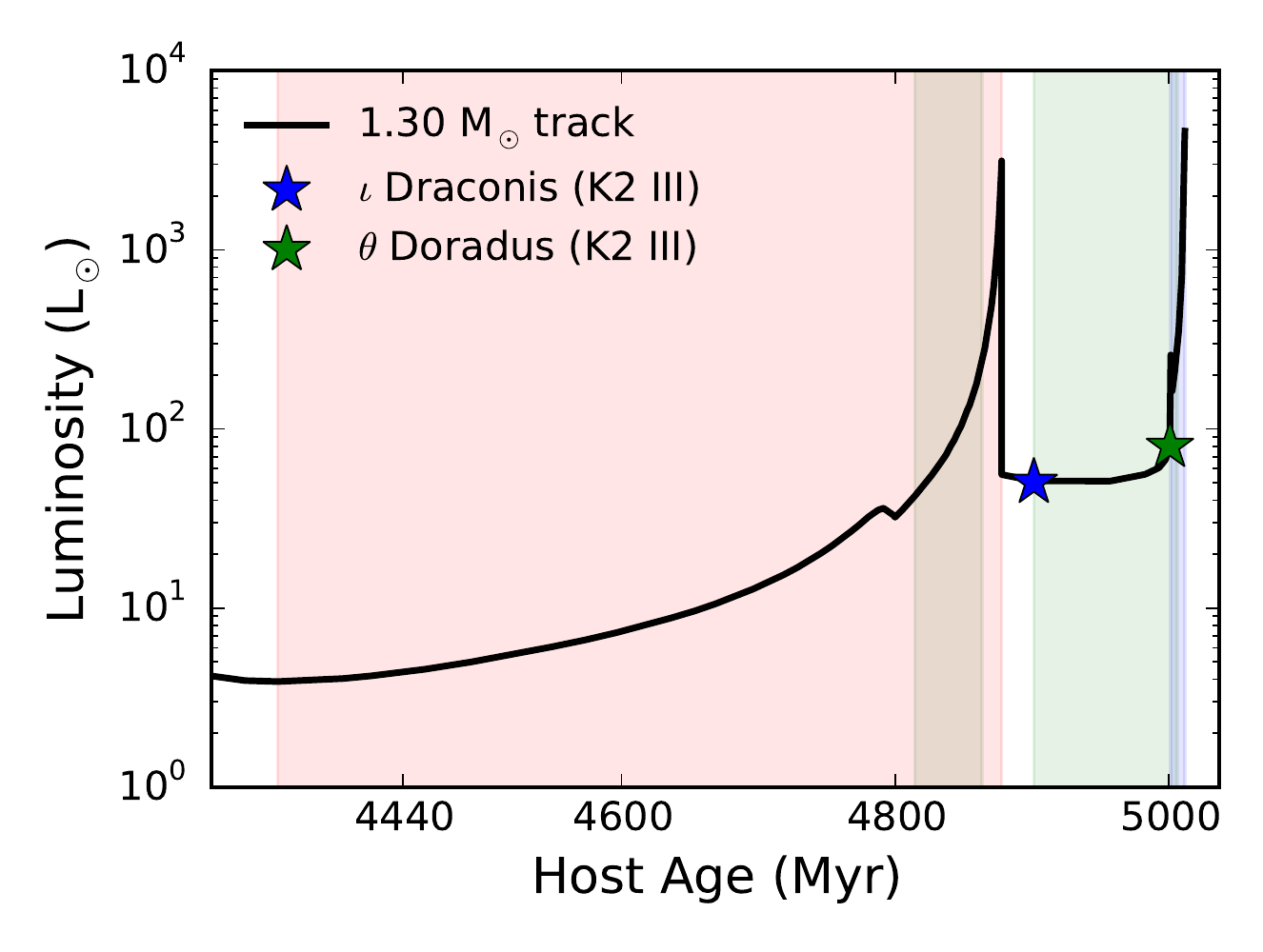}
\includegraphics[scale=0.57]{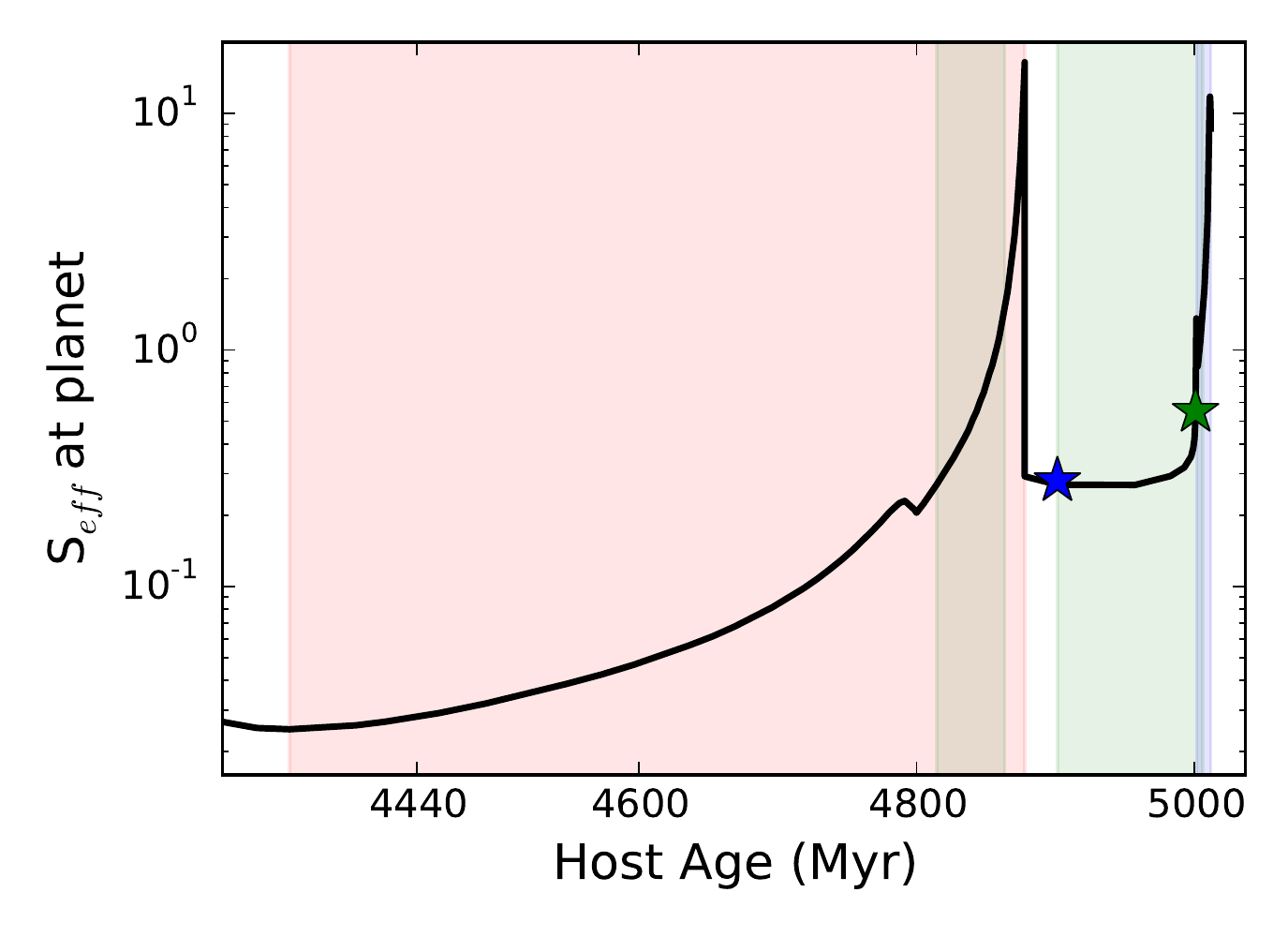}\\
\includegraphics[scale=0.57]{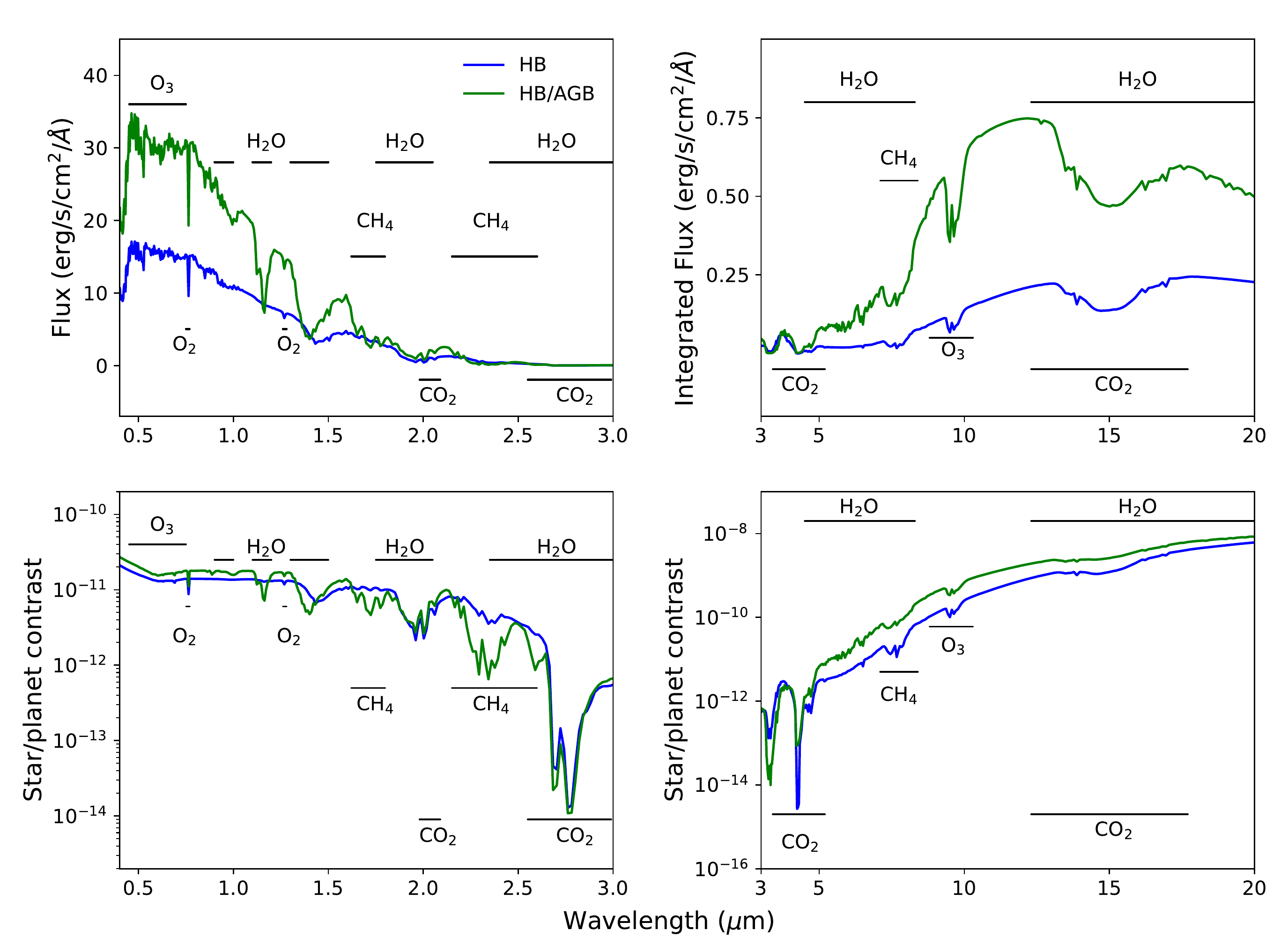}
\caption{Luminosity evolution and spectra of a planet in the HZ of a 1.3 M$_\odot$ red giant on the orbit giving the maximum amount of time in the RG HZ. The top row shows (left) the luminosity evolution of the red giant host and (right) the changing incident flux upon the planet on an orbit giving it maximum time in the habitable zone. The quantity S$_{\footnotesize \mbox{eff}}$ is the fraction of flux received by the planet compared to modern Earth. The modeled points of evolution are marked with stars. The middle and bottom rows show the spectra of this planet throughout evolution and the star/planet contrast ratios.  \label{HZ_tracks}}
\end{figure}

Figure~\ref{HZ_tracks} shows the evolution of a model planet's spectra set at an initial orbital distance of 12.5 AU from a 1.3 M$_\odot$ host star during its RG evolution in planetary flux and planet-to-star contrast ratio as an example for the 3 RG mass tracks modeled. The points in time modeled are shown as star markers on the RG's evolutionary track in the top row. The red zone indicates the RG phase, the blue zone the AGB, and the green zone the time that the specific model planet spends in the HZ during the RG's evolution.

These planets spend the majority of their time continuously in the RG HZ initially orbit near the outer edge of the RG HZ where the overall incident irradiation is low (Figure~\ref{HZ_tracks} top right). As a result these planet models initially have cold surface temperatures because we did not change the CO$_2$ concentration in these models for ease of comparison. Assuming rocky planets in the HZ of RGs could maintain a carbonate silicate cycle until the RG phase of their host, these planets should maintain higher CO$_2$ concentration in their atmosphere (see e.g.\ review \citealt{kalt17}). However there are no models which explore whether the carbonate-silicate or similar cycles could be maintained on frozen planets until the RG phase of their host. Therefore we maintained a modern Earth-analog concentration of CO$_2$ in our planet models, which initially produce cold environments until the RGs luminosity increases sufficiently to warm the planet.

Cold planetary atmospheres generally have lower H$_2$O abundance, resulting in initially shallower H$_2$O spectral features, which become more pronounced with a RGs increasing luminosity and the resulting increase in planetary surface temperature. In the reflected spectra the increase is due to the increasing H$_2$O abundance, while in the emission spectra the increasing H$_2$O abundance as well as the increasing planet emission temperature deepen the absorption features. Increasing UV incident flux on the planets over the RG's evolution also increases the O$_3$ abundance and decreases the CH$_4$ abundance in the planets' atmospheres (see Figure~\ref{HZ_tracks} middle). 

These model planets initially have low amounts of reflected flux due to the smaller incident amounts of RG host flux at their larger orbital distance and lower emitted flux due to the low planetary surface temperatures at the outer edge of the RG HZ. With increasing incident irradiation at the planet's orbital position from the RG through its evolution (Figure~\ref{HZ_tracks} top right), the planet's reflected flux increases. The planet's emitted flux also increases due to its increasing surface temperature. Thus a planet in the HZ of a RG displays increasing depth of spectral features in both emitted and reflected flux with its RG's evolution.

Figure~\ref{HZ_tracks} (bottom) shows that the star to planet contrast ratio stays similar over the RG's evolution for planets at a certain distance within in time they remain in the RG HZ limits, with a slight decrease with the hosts evolution.

\subsection{Estimated observation times for red giant HZ planets}

\begin{figure}[h!]
\centering

\includegraphics[scale=0.5]{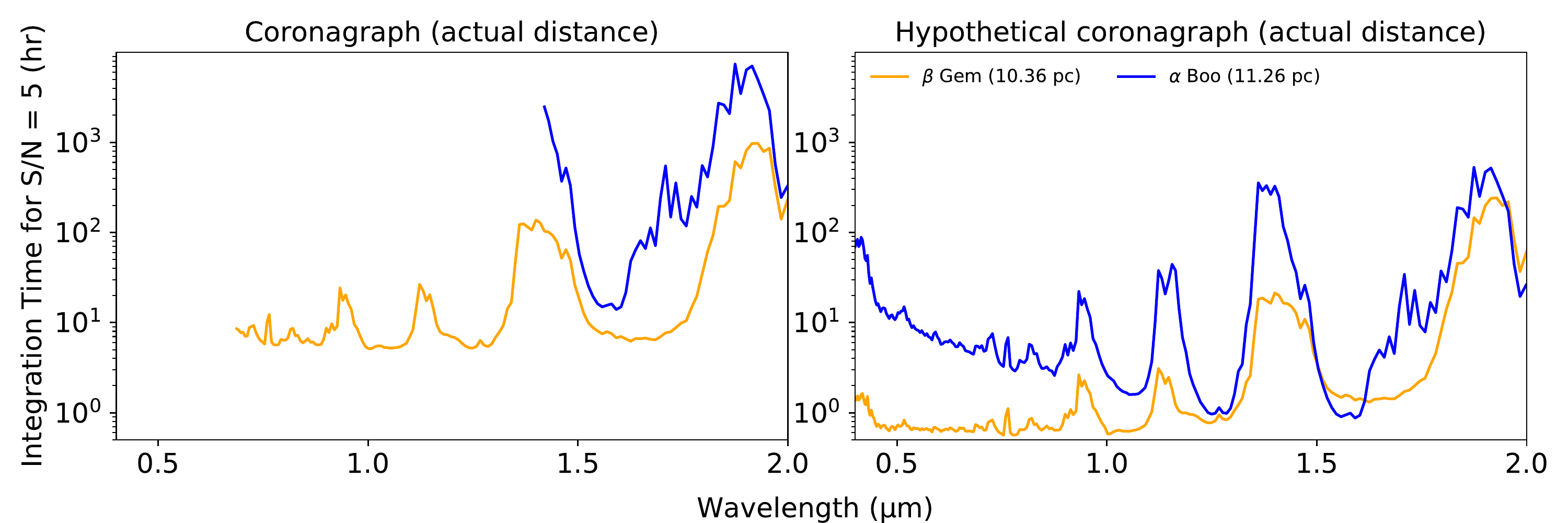}\\
\includegraphics[scale=0.5]{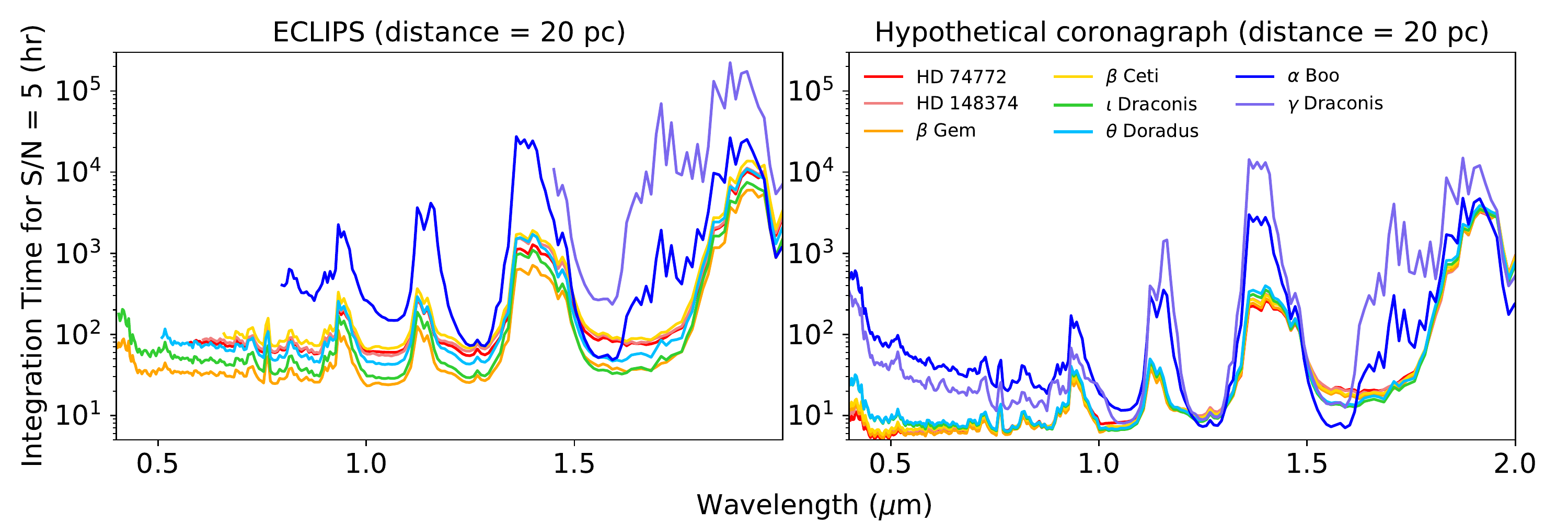}
\caption{Integration times necessary to reach a signal-to-noise ratio of 5 for the 15 m space-based LUVOIR-A concept with (left) the predicted parameters of the ECLIPS coronagraph, and (right) a hypothetical optimized coronagraph with a $10^{12}$ light suppression at the RG HZ and an OWA of 230 $\lambda/D$, for nominal RG HZ planets. Plots on the top row show the two closest RGs to Earth, $\beta$ Gem and $\alpha$ Boo, at their actual distances of 10.26~pc and 11.36~pc, respectively. The bottom row shows all RG hosts at a standardized distances of 20 pc to compare spectral feature and integration time differences. Wavelengths with no information are due to the system being outside of the OWA for ECLIPS.  Longer wavelengths are not shown as the low amount of flux from the star results in very small reflected light fluxes.  \label{luvoir}}
\end{figure}

While no existing or proposed instrument concept has been designed to observe RG HZ planets, as an example of how such planets could be observed, we use one of the proposed coronagraphs of the mission concept, LUVOIR-A, a 15~m diameter space-based telescope concept, to estimate the required integration times in the visible/NIR to observe spectral features in reflected light for RG HZ planets. Using a coronagraph simulator \citep{robi16,lust19} we estimate integration times required to reach a SNR of 5 for our model planets using i) LUVOIR's ECLIPS coronagraph design, with a proposed suppression of 10$^{10}$ between the IWA of 3.5 $\lambda/D$ and the OWA of 64 $\lambda/D$, and ii) an imagined hypothetical coronagraph optimized for RG HZ planets, which allows for a larger suppression of 10$^{12}$ at the wider orbital separation of the RG HZ, with an OWA of 230 $\lambda/D$. Figure~\ref{luvoir} shows the results of these simulations for the two closest RGs to Earth, and for all RG hosts used in this paper normalized to a standard distance of 20~pc.  About 20 RGs can be found within 30~pc of the Sun (see Table~10 in \citealt{koza19}).

Spectra of model planets orbiting cooler RGs generally need longer integration times due to the smaller star-planet contrast ratios (see Figure~\ref{fig:RG_spectra}).  This can be seen comparing integration times of our two closest RGs (seen in the top row of Figure~\ref{luvoir}), $\beta$ Gem (with the highest contrast ratios of our targets) and $\alpha$ Boo (with the second lowest contrast ratio).   For example, to reach a SNR of 5 for the 0.76 \mm\ O$_2$ line with $\beta$ Gem it would require 12.4 hours of observations with ECLIPS, versus 1.1 hours with our hypothetical optimized coronagraph.  For $\alpha$ Boo, located at roughly the same distance, the 0.76 \mm\ line cannot be observed with ECLIPS due to the restricting OWA, and would require 6.8 hours of observation with our hypothetical coronagraph, due to the worse contrast ratio.

Assuming that the increased orbital distance of the RG HZ would allow for a increase in light suppression, our envisioned hypothetical, optimized coronagraph, with a 10$^{12}$ suppression at the RG HZ and a large OWA of 230 $\lambda/D$, would allow much shorter integration times than with LUVOIR's ECLIPS.  Such a coronagraph would be necessary to effectively characterize RG HZ planets.

\section{Discussion and Conclusions \label{conclusions}}

While attention is focused on planets orbiting in the HZ of MS stars, detected exoplanets orbiting RG hosts raise the question of the existence of life on initially frozen worlds, persistence of life and how to characterize it on such planets.
It is unknown to date, what kind of life could develop and thrive on initially frozen worlds and how such life could evolve during the RG phase of its host on a warming planet. Therefore, as a starting point to explore how to search for life around RG stellar hosts, we simulated Earth-like planets around a variety of RG hosts \citep{koza19}, both orbiting i) at the distance where they would receive similar irradiation as modern Earth, as well as ii) at a distance which allows for the longest continuous time in the RG HZ.

The high resolution reflection and emission spectra from 0.4 to 20 \mm\ for models of nominal Earth-like planets orbiting in the HZ of RGs (plotted at a resolution of 140 for clarity in Figures~\ref{fig:RG_spectra}) show atmosphere features of H$_2$O, CO$_2$, CH$_4$, O$_3$ and O$_2$. Identification of all spectral features improve with resolution, as shown for one example, O$_2$ at 0.76 \mm\ in reflected light (plotted at a resolution of 100,000 Figure~\ref{fig:O2} as proposed for spectrographs like HIRES at ELTs). 

Our noise-free high resolution spectra (minimum resolution 100,000) are available online\footnote{https://doi.org/10.5281/zenodo.4027344} to serve as input for simulations of observations for ground- and space-based telescopes. Observation time for specific features will dependent in addition on e.g.\ stellar noise and properties and systematics of the observing instrument. 

The orbital distance of the RG HZ increases significantly due to the higher luminosity of the host during its RG phase, increasing the apparent angular separation of the RG HZ significantly compared to the MS HZ. However, the contrast ratio between RG hosts and such nominal HZ planets also increases by up to 3 orders of magnitude compared to similar planets orbiting in the HZ of MS stars. 

No existing or proposed instrument concept has been envisioned yet to observe terrestrial planets in the RG HZ. Such planets  would require larger observation times than planets in the HZ of MS stars, due their RG hosts' increased luminosity. We modeled necessary integration times to reach a SNR of 5   using the  proposed future mission concept LUVOIR to quantify the detectability of such planets (see Figure~\ref{luvoir}).  Using predicted parameters of LUVOIR's ECLIPS coronagraph we found that the higher luminosity of RG hosts led to long integration times, and that the larger separation of the RG HZ compared to MS star HZs resulted in spectral cutoffs due to ECLIPS's OWA.  Assuming that the increased orbital distance of the RG HZ could allow for a increase in light suppression at that distance, we envision a hypothetical, optimized coronagraph, with a 10$^{12}$ suppression at the RG HZ with a larger OWA, which would reduce the observation times considerably, making terrestrial planet in the HZ of RG interesting targets to observe.

Ground-based high resolution spectroscopy has already characterized atmospheric species for unresolved planets like HD 179949 b using the planets known motion during the observations.We extrapolate that a similar approach could be used to counter the increased luminosity of RG hosts to observe their HZ planets, which would make such planets interesting targets for the extremely large telescopes like the ELT, the Giant Magellan Telescope (GMT), and the Thirty Meter Telescope (TMT).

Our high resolution spectral database of nominal habitable worlds orbiting evolved RG hosts can be used to explore possibilities for future observations of such planets to assess whether their atmospheres can give us first insights into whether life could survive and thrive on such formerly icy worlds.

\acknowledgements

This work was supported by the Carl Sagan Institute. T.K.\ was additionally funded by a NASA Space Grant fellowship.

\end{document}